\documentclass[12pt]{iopart}

\usepackage{epsfig}

\def\8{\infty}
\def\oh{\frac{1}{2}}

\def\d{\partial}

\def\undertext#1{\vtop{\hbox{#1}\kern 1pt \hrule}}

\def\VEV#1{\left\langle\,#1\,\right\rangle}

\def\pp#1{\frac{\partial}{\partial#1}}
\def\pbyp#1#2{\frac{\partial#1}{\partial#2}}

\def\be{\begin{equation}}
\def\ee{\end{equation}}
\def\bea{\begin{eqnarray} & &}
\def\eea{\end{eqnarray}}

\def\rf#1{(\ref{#1})}

\def\rfs#1{Eq.~\rf{#1}}

\begin{document}

\title[Logarithmic operators and logarithmic conformal field theories]{Logarithmic operators and logarithmic conformal field theories}

\author{Victor Gurarie}

\address{Department of Physics CB390,
University of Colorado,
Boulder, CO, 80309}
\begin{abstract}
Logarithmic operators and logarithmic conformal field theories are reviewed. Prominent examples considered here include  $c=-2$ and $c=0$ logarithmic conformal field theories. $c=0$ logarithmic conformal field theories are especially interesting since they describe some of the critical points of a variety of longstanding problems involving a two dimensional quantum particle moving in a spatially random potential,  as well as critical two dimensional  self avoiding random walks and percolation. Lack of classification of logarithmic conformal field theories remains a major impediment to progress towards finding 
complete solutions to these problems. 
 \end{abstract}

\maketitle

\section{Introduction}
\label{sec:intro}
The birth of conformal field theory can rightfully be considered to be in 1984 when the celebrated paper by A. Belavin, A. Polyakov and A. Zamolodchikov was published \cite{Zamolodchikov1984}. Despite being extraordinary comprehensive, one natural possibility was not considered at all in that paper, the possibility that correlation functions of a conformally invariant quantum field theory may contain logarithms. It took some time to discover that the logarithms do appear in some conformal field theories \cite{Saleur1992}, and they do it without breaking scale and conformal invariance despite what one might fear. Finally, in 1993 Ref.~\cite{Gurarie1993}
introduced the concept of logarithmic operators as the ones responsible for the logarithms in the correlators in conformal field theory while preserving conformal invariance. 

It is probably fair to say that just a few people were aware of the logarithmic operators until 1996 when several publications firmly put the subject on the map. They include, but are not limited to, Refs.~\cite{Tsvelik1996,Kausch1996,Kausch1996b,Flohr1996}. After that, the theory of logarithmic operators underwent rapid development in many directions. Nevertheless, logarithmic conformal theories remain very difficult to work with, and are not yet classified in a way similar to the standard conformal field theories which contain no logarithms. 

It is particularly frustrating that certain types of problems in condensed matter physics, the ones involving quantum motion in the presence of a spatially random potential \cite{EfetovBook}, map at their critical points to conformal field theories at central charge $c=0$ (see for example the discussion in Ref.~\cite{Gurarie1999} and references therein). In turn these conformal field theories are expected not only to possess logarithmic operators corresponding to  various physical observables \cite{Maassarani1997,Cardy1999,Gurarie2002} but also include logarithmic operators extending the Virasoro algebra itself \cite{Gurarie2002,KoganBook}. Among the problems such conformal field theories are presumed to describe is the famous quantum Hall plateau transition, a localization transition in the motion of a quantum particle in two dimensions in a uniform magnetic field and in a spatially random potential (see Ref.~\cite{Huckenstein1995} and references therein). The critical properties of this transition are subject to much discussion in the literature, with the critical exponent initially conjectured to be close to $\nu \approx 7/3$, with some even suggesting that
this number is larger exactly by $1$ \cite{Sokolov1988} than the known percolation critical exponent $\nu_{\rm perc} = 4/3$ \cite{Dennijs1979}, but more recently found numerically to have a higher value, at least $\nu \approx 2.5$ \cite{Amado2009} or even as high as $\nu \approx 2.6$ \cite{Slevin2009}. It is now believed that the solution to the problem of exact computation of this and similar critical exponents via conformal field theory must involve logarithms in some crucial way, and that the main impediment to finding this solution is due to the fact that logarithmic conformal field theories have not yet been systematically classified.

It is also interesting to observe that two dimensional percolation and self-avoiding random walks have been solved exactly via  maps to the $Q \rightarrow 1$ limit of the $Q$-state Potts model and the $N \rightarrow 0$ limit of the O$(N)$ models respectively (the maps are described, for example, in Ref.~\cite{CardyBook}). These maps proceed via constructing the exact solution of $Q>1$ and $N>0$ models with the help of conformal field theory at $c>0$, and then taking the appropriate limit which corresponds to taking $c \rightarrow 0$ in the exact solution. Interestingly, while the conformal field theory at $c>0$ contains no logarithms and is largely understood, and while the critical exponents can be extracted by taking the $c\rightarrow 0$ limit in the dimensions of the appropriate operators, the limiting conformal field theories describing percolation and self-avoiding random walks at $c=0$ contain logarithms and have not yet been completely understood despite the availability of the limiting procedure. In fact, there is a lot of similarity between the conformal field theories describing percolation and self-avoiding random walks and the ones describing critical points with quenched disorder (meaning, involving a time-independent random in space variable), and it is believed that understanding the former will help understanding the latter \cite{Saleur2012}. 

The goal of this paper is largely to review the contributions of the author and collaborators to the study of  logarithmic conformal field theory, mostly contained in Refs.~\cite{Gurarie1993,Gurarie1997,Gurarie2002,KoganBook}. This paper is not a comprehensive review of the subject, but hopefully it will contribute to stimulating research in this potentially exciting direction. 

\section{Logarithmic operators}
\label{sec:log}
Logarithmic operators are a straightforward generalization of the concept of  primary operators. Primary states (states created by the primary operators) are annihilated by all the Virasoro generators $L_n$ with $n>0$ and are
eigenstates of the Virasoro generator $L_0$,
\be L_0 \left| A \right> = h \left| A \right>.
\ee
Logarithmic operators are a generalization of that to non-diagonalizable matrices. The logarithmic states are also annihilated by $L_n$ with $n>0$, but form a Jordan block with respect to $L_0$,
\be \label{eq:logop}  L_0 \left| C \right> = h \left| C \right>, \ L_0 \left| D \right> = h \left| D \right> + \left| C \right>, \ee or equivalently
\be \label{eq:jordan} L_0
\left( \begin{matrix} { \left| C \right> \cr \left| D \right>  } \end{matrix} \right) =  \left( \begin{matrix} { h & 0 \cr 1 & h  } \end{matrix} \right) \left( \begin{matrix} { \left| C \right> \cr \left| D \right>  } \end{matrix} \right).
\ee
\rfs{eq:jordan} is an example of a  non-diagonalizable Jordan block. Obviously  Jordan blocks of an arbitrary size are also possible, but for the sake of simplicity let us restrict our attention to the two by two block as in \rfs{eq:jordan}. We see from \rfs{eq:logop} that $\left| C \right>$ looks like a regular primary state. 
At the same time, the result of applying $L_0$ to  $\left| D \right>$ is somewhat different and involves $\left| C \right>$ as well as $\left| D \right>$. $\left|D \right>$ can be termed a logarithmic partner of $\left| C \right>$, for reasons which will become obvious  in a moment. 

Note that $L_0$ can be interpreted as the Hamiltonian since it generates time translations in the usual radial quantization of conformal field theory. It follows from \rfs{eq:logop} that the Hamiltonian is not hermitian since $L_0^\dagger \not = L_0$. Therefore, conformal field theories with logarithmic states cannot be unitary.

In conformal field theory there is an operator which corresponds to every state, so there must exist operators $C(z)$ and $D(z)$. It is straightforward to calculate their two-point correlation functions by conformal invariance. Under an infinitesimal conformal transformation given by $w=z+\epsilon(z)$, a primary operator transforms as
\be \delta C(z) = \epsilon(z) \pbyp{C(z)}{z} + h \pbyp{\epsilon(z)}{z} C(z).
\ee
Its logarithmic partner $D(z)$ obviously  transforms in a slightly different way
\be \delta D(z) = \epsilon(z) \pbyp{D(z)}{z} +  \pbyp{\epsilon(z)}{z} \left( h D(z) + C(z) \right).
\ee One of the basic facts of conformal field theory states that 
demanding that the two point correlation function of a primary operator $A(z)$ is invariant under translations  where $\epsilon(z)=\epsilon={\rm const}$, dilatations with $\epsilon(z) = \epsilon z$ and special conformal transformations (SCT) with $\epsilon(z)=\epsilon z^2$, fixes that correlation function up to a constant
\be \label{eq:twopoint} \VEV{A(z_1) \, A(z_2)} = \frac{B}{\left( z_1- z_2 \right)^{2h}},
\ee
where $B$ is an arbitrary constant. 
This can be established by writing down the differential equations satisfied by this correlation function
\begin{eqnarray} \label{eq:inv1} {\rm translations:} && \left( \pp{z_1} + \pp{z_2} \right) \VEV{A(z_1) \, A(z_2)} = 0, \cr
{\rm dilatations:} && 
 \left( z_1 \pp{z_1} + z_2 \pp{z_2} +2 h \right) \VEV{A(z_1) \, A(z_2)} =0, \cr
{\rm SCT:} &&  \left( z_1^2 \pp{z_1} + z_2^2 \pp{z_2} +2 h \left(z_1+z_2 \right) \right)  \VEV{A(z_1) \, A(z_2)} =0.
\end{eqnarray}
As is well known, \rfs{eq:twopoint} is the only solution to these equations, up to the arbitrary constant $B$.

A correlation function $\VEV{C(z_1) C(z_2)}$ obviously satisfies the same set of equations, with the same result as \rfs{eq:twopoint}. A similar procedure for the two point functions involving $D(z)$ leads to the
following equations, slightly modified compared to those satisfied by $\VEV{C(z_1) \, C(z_2)}$, 
\begin{eqnarray}  \label{eq:inv} {\rm translations:} &&  \left( \pp{z_1} + \pp{z_2} \right) \VEV{D(z_1) \, C(z_2)} = 0, \cr &&  \left( \pp{z_1} + \pp{z_2} \right) \VEV{C(z_1) \, D(z_2)}=0, \\
&&  \left( \pp{z_1} + \pp{z_2} \right) \VEV{D(z_1) \, D(z_2)}=0; \cr
{\rm dilatations:} &&  \left( z_1 \pp{z_1} + z_2 \pp{z_2} +2 h \right) \VEV{D(z_1) \, C(z_2)} +  \VEV{C(z_1) \, C(z_2)}= 0, \cr  &&  \left( z_1 \pp{z_1} + z_2 \pp{z_2} +2 h \right) \VEV{C(z_1) \, D(z_2)} +  \VEV{C(z_1) \, C(z_2)}= 0, \cr &&  \left( z_1 \pp{z_1} + z_2 \pp{z_2} +2 h \right) \VEV{D(z_1) \, D(z_2)} + \cr && + \VEV{C(z_1) \, D(z_2)} + \VEV{D(z_1) \, C(z_2)}= 0; \cr
{\rm SCT:} &&  \left( z_1^2 \pp{z_1} + z_2^2 \pp{z_2} +2 h \left(z_1+z_2 \right) \right) \VEV{D(z_1) \, C(z_2)} + \cr && +  2  z_1 \VEV{C(z_1) \, C(z_2)}= 0, \cr &&  \left( z_1^2 \pp{z_1} + z_2^2 \pp{z_2} +2 h \left( z_1+z_2 \right) \right) \VEV{C(z_1) \, D(z_2)} \cr && + 2  z_2 \VEV{C(z_1) \, C(z_2)}= 0, \cr
&&  \left( z_1^2 \pp{z_1} + z_2^2 \pp{z_2} +2 h \left( z_1+z_2 \right) \right) \VEV{D(z_1) \, D(z_2)} + \cr && + 2  z_1 \VEV{C(z_1) \, D(z_2)}+2  z_2 \VEV{D(z_1) \, C(z_2)}= 0. \nonumber
\end{eqnarray}
The only solution of these equations states that
\begin{eqnarray} \label{eq:logcorr} && \VEV{C(z_1) \,  C(z_2)} = 0, \  \VEV{D(z_1)\, C(z_2)} = \VEV{C(z_1) \, D(z_2)} = \frac{B}{(z_1-z_2)^{2h}}, \cr && \VEV{D(z_1) \, D(z_2)} = -2 B\frac{\ln (z_1-z_2)}{(z_1-z_2)^{2h}}.
\end{eqnarray}
As before, here $B$ is an arbitrary constant which cannot be fixed by conformal invariance alone. 

We see that the operator $D(z)$ justifies its name of a logarithmic operator, as its two point correlation function  contains a logarithm.  It should be emphasized that despite the logarithm, the correlation functions of $D(z)$ are fully scale and conformally invariant. 

We also see that the correlation function $\VEV{C(z_1) \, C(z_2)}$ is zero. One  can show that if it is not zero, then  \rfs{eq:inv} cannot be satisfied. Thus the primary partner of the logarithmic operator must have vanishing norm. 

We see that logarithmic operators naturally generalize the concept of a primary operator. What is also interesting is that they routinely occur  in the theories whose primary operators belong to the Kac table \cite{Zamolodchikov1984}, that is, their correlation functions satisfy differential equations. 

A primary operator $A(z)$ with a conformal dimension $h$  has a four point function which is generally given by
\be \VEV{A(z_1) \, A(z_2) \, A(z_3) \, A(z_4) } = \frac{1}{\left( z_1- z_3 \right)^{2h} \left( z_2-z_4 \right)^{2 h} } F(x),
\ee
where $x$ is given by
\be x = \frac{(z_1-z_2)(z_3-z_4)}{(z_1-z_3)(z_2-z_4)}.
\ee
If the operator $A(z)$ belongs to the Kac table, the function $F(x)$ satisfies an ordinary differential equation.  The solution of this equation has singularities at $x=0$, $x=1$, and $x=\infty$, and at those singularities typically $F(x)$ behaves as a power law. That power law is a reflection of the underlying operator product expansion (OPE) of the operator $A(z)$ with itself. 

For example, if $F(x) \sim x^{\alpha}$ for $x \rightarrow 0$, then it is straightforward to see that the OPE of $A(z)$ with itself has the leading term
\be A(z) A(0) \sim z^\alpha B(0) + \dots,
\ee where $B$ is some other primary operator with the dimension $\delta=\alpha+2h$. 

However, in some cases the differential equations may have logarithmic singularities instead of the more ordinary power laws at those special points. Suppose the leading singularity of $F(x)$ as $x \rightarrow 0$ is
\be F(x) \sim x^{\alpha} \ln x + \dots.
\ee
Then the only explanation of this behavior in terms of the OPE is that
\be \label{eq:opel}  A(z) A(0) \sim z^{\alpha} \left( C(0) \ln z + D(0) \right) +\dots. 
\ee
Here $D$ is the logarithmic partner of $C$ as introduced above. Notice that this is consistent  with and indeed requires that $C(z)$ has  a vanishing two point function with itself. If its correlation function with itself had not been  zero, then $F(x)$ would have necessarily behaved as $\ln^2 x$. 

The OPE \rfs{eq:opel} is conformally invariant itself. That can be checked by using the standard commutation relation
\be \label{eq:st} \left[ L_n, A(z) \right] = z^{n+1} \pbyp{A(z)}{z} +h   (n+1) z^n A(z).
\ee
Applying $L_0$ to $A(z) A(0)$ gives
\begin{eqnarray} L_0 A(z) A(0) & =& \left[ L_0, A(z) \right] A(0) + A(z) L_0 A(0) = \cr &=&  \pbyp{A(z)}{z} A(0) +2h    A(z) A(0).
\end{eqnarray}
Substituting \rfs{eq:opel} gives
\be \label{eq:lhs} \delta z^\alpha  \left( C(0) \ln z + D(0) \right) + z^\alpha C(0),
\ee
where $\delta = \alpha + 2h $ is the dimension of $C(z)$. At the same time,
applying $L_0$ directly to the right hand side of \rfs{eq:opel} and using that in this context
\be L_0 \left| C \right> = \delta \left| C \right>, \ L_0 \left| D \right> = \delta \left| D \right> + \left| C \right>,
\ee
we obtain
\be \label{eq:last} z^\alpha  L_0 \left( C(0) \ln z + D(0) \right) = \delta z^\alpha  \left( C(0) \ln z + D(0) \right) + z^\alpha C(0)
\ee which coincides with \rfs{eq:lhs}. 

A situation where $F(x)$ has logarithms typically arises whenever the fusion rules of $A(z)$ with itself (or with some other operator from the Kac table), computed according to the rules of the Kac table, produce several operators on the right hand side with coinciding dimensions, or with dimensions different by an integer. These coinciding dimensions is what causes logarithms. We will see the examples of this in the next section. 

Needless to say, this never occurs in unitary minimal models since if it occurred it would violate unitarity. However, this routinely occurs in the non-unitary minimal models. 

Thus we discovered that logarithmic operators are self-consistent, have conformally invariant correlation functions, and appear on the right hand side of the OPE of the usual primary operators if those primary operators belong to the Kac table, and if the corresponding four point functions have logarithmic singularities as a consequence of the differential equations they satisfy. 

\section{Conformal field theory at $c=-2$ and zero dimensional logarithmic operators}
\label{sec:minustwo}
An especially simple example of a logarithmic operator occurs in the theory with the central charge $c=-2$. This theory is represented by a functional integral
\be \label{eq:poaction} Z_f = \int {\cal D} \psi {\cal D} \bar \psi \, e^{ - \int d^2 z \, \d \psi \bar \d \bar \psi}.
\ee
Here $\psi$ and $\bar \psi$ are fermionic (anticommuting) variables. 

The fact that this critical  theory contains logarithms was pointed out in Ref.~\cite{Gurarie1993}, and subsequently this was a subject of intense research (see, for example, Ref.~\cite{Kausch2000}). Let us go over
the arguments why this occurs in this theory.

This theory appears similar to the analogous theory of complex boson
\be Z_b= \int {\cal D} \phi {\cal D} \bar \phi \, e^{- \int d^2 z \bar \d \bar \phi \d \phi}
\ee
where $\phi$ is a complex commuting field, but it also has crucial differences. On the one hand,
the stress energy tensor of both theories appear to have the same form
\be T_f  \sim  \d \bar \psi \d \psi, \ T_b \sim  \d \bar \phi \d \phi.
\ee
On the other hand, combined with the OPE 
\be \d \bar \psi(z) \d \psi(0) \sim \frac{1}{z^2} + \dots, \  \d \bar \phi(z) \d \phi(0) \sim \frac{1}{z^2} + \dots
\ee
which also appears basically the same in both theories, one verifies that the central charge of the
fermionic theory is $c=-2$ unlike the central charge of the bosonic theory which is $c=+2$ (the central charge of a free complex
boson). 

More differences appear upon further analysis of the fermionic theory, which we present here  following the discussion in Ref.~\cite{Gurarie1997}. The fermionic variable may contain zero modes, constant pieces in the fields $\psi$ and $\bar \psi$, which do not enter the action in 
\rfs{eq:poaction}. Integrating over this mode, in accordance with the rules of Grassmanian variable integration, gives zero for $Z_f$. More formally, we should split
\be \psi(z)= \xi + \zeta(z), \ \bar \psi(z) = \bar \xi + \bar \zeta(z),
\ee
where the constant pieces $\xi$ and $\bar \xi$  in $\psi(z)$ and $\bar \psi(z)$ were explicitly indicated. Then the functional integral should really be understood as
\be Z_f= \int d\xi d\bar \xi  \, {\cal D} \zeta {\cal D} \bar \zeta \, e^{-\int d^2 z \, \d \zeta \bar \d \bar \zeta}.
\ee
With this clarification of the meaning of the functional integral in \rfs{eq:poaction}, we find
that 
the expectation value of the vacuum is just 0, \be \left<  I \right>= \int d\xi d\bar \xi  \, {\cal D} \zeta {\cal D} \bar \zeta \, e^{-\int d^2 z \, \d \zeta \bar \d \bar \zeta}=0,\ee 
thanks to the integration over Grassmanian $\xi$ and $\bar \xi$. 
Here $I$ is the identity operator. At the same time, the expectation value of the object $: \bar \psi \psi:$, where the colon indicates normal ordering, is a nonzero constant,
which we will call $B$, 
\be \left<  : \bar \psi \psi  :\right> = \int d\xi d\bar \xi  \, {\cal D} \zeta {\cal D} \bar \zeta \, \bar \xi  \xi \, e^{-\int d^2 z \, \d \zeta \bar \d \bar \zeta} =B.\ee
It is tempting to identify the identity $I$ with the operator $C$ and $: \bar \psi \psi:$ with $D$, in this case both having dimension $h=0$. 
Further evidence that this is the correct identification comes from the observation that
\be \VEV{ : \bar \psi(z_1) \psi(z_1) : \, : \bar \psi(z_2) \psi(z_2):} = - 2 B \ln \left( z_1-z_2 \right).
\ee
This correlation function can be found by decomposing $\psi$ into the sum of $\xi$ and $\zeta$, and then doing the functional integrations over $\xi$ and $\zeta$. 

We see that the identifying $: \bar \psi \psi$ with the logarithmic partner of the identity operator is consistent, as it produces the correct correlation function in agreement with \rfs{eq:logcorr} where $h$ is set to zero. 

One might think that one never has to deal with this logarithmic operator if one restricts the attention to the simpler conformal operators $\d \psi(z)$ and $\d \bar \psi(z)$. However, it is not quite true either. Imagine we consider a situation where the fields $\psi(z)$ and $\bar \psi(z)$ are antiperiodic as one goes around the origin of the complex plane. It is easy then to convince oneself that 
\be \label{eq:anti} \VEV{ \partial \psi(z) \,  \partial \bar \psi(w)} \sim \frac{\frac 1 2 \left( \sqrt{\frac z  w} + \sqrt{\frac w z} \right)}{(z-w)^2}.
\ee
This correlation function has a singularity as $w \rightarrow 0$ or $z \rightarrow 0$. Therefore, the vacuum here is no longer trivial, rather it is created by an operator $\sigma(z)$ whose dimension can be read off \rfs{eq:anti}. A straightforward calculation involving taking the limit $z \rightarrow w$ in \rfs{eq:anti} and identifying the resulting correlator with the expectation of the stress-energy tensor, shows that this dimension is $-1/8$.  The operator $\sigma(z)$ plays a role not unlike the 
order operator of the Ising model in the theory of $c=1/2$ Majorana fermions (imposing antiperiodic boundary conditions on the Majorana fermions produces an operator with dimension $1/16$, the order
operator of the Ising model). 

A question comes up then, if $\sigma$ is an unavoidable field if the correlation functions of $\d \psi$ are considered, what are the correlation functions of $\sigma$? To answer that, one needs to use the
fact that $\sigma$ is a degenerate primary operator, belonging to the Kac table at the central charge $c=-2$. Let us recall that for the central charges satisfying
\be c = 1- 6 \frac{\left( p - q\right)^2}{p q},
\ee for nonnegative integers $p$ and $q$, there is a set of degenerate operators with dimensions
\be \Delta_{p,q} = \frac{\left( n p - m q \right)^2 -\left( p - q\right)^2}{4 p q},
\ee whose correlation functions satisfy differential equations of the $p \cdot q$ order. 

For example, if $p=2$ and $q=1$, $c=-2$. The table of the degenerate operators (called the Kac table) then takes the following form
\begin{center} \begin{tabular}{  l || c | c | c| c| r}  $n$ \textbackslash \, $m$ & 1 & 2 & 3 &4 & \dots \\ \hline \hline  \noalign{\smallskip} 1 & 0 & $-\frac 1 8$  & 0 & $\frac 3 8$ & \dots \\ 
\noalign{\smallskip} \hline \noalign{\smallskip} 2 & 1 & $\frac 3 8$  & 0 & $-\frac 1 8$ & \dots \\
\noalign{\smallskip} \hline \noalign{\smallskip} 3 & 3 & $\frac {15} 8$  & 1 & $\frac 3 8$ & \dots \\
\noalign{\smallskip} \hline \noalign{\smallskip} \dots & \dots & \dots & \dots & \dots & \dots
\end{tabular} 
\end{center}
For example, the operator with the position $(n,m)=(2,1)$ in this table has dimension $1$ and can be identified with the operators $\d \psi$ and $\d \bar \psi$. Indeed, with some work one
can check that the correlation functions of these operators which can be computed directly by definition using the Gaussian functional integral \rfs{eq:poaction} satisfy the appropriate differential equations. 

The operator with the position $(1,2)$ in this table has dimensions $-1/8$ and can be identified with the operator $\sigma(z)$ introduced above. One can now study its correlation functions
by solving the appropriate differential equation. One can verify that the solution of this equation gives
\be \label{eq:fourpoint} \VEV{\sigma(z_1) \, \sigma(z_2) \, \sigma(z_3) \, \sigma(z_4)} = \left[ \left(z_1-z_3 \right) \left( z_2 - z_4 \right) x (1-x) \right]^{\frac 1 4}F(x).
\ee
Here $F(x)$ depends on the choice of the conformal block, and is equal to either the hypergeometric function $F_1(x)=F\left( \frac 1 2, \frac 1 2, 1 ; x\right)$, or the function $F_2(x)=F\left( \frac 1 2, \frac 1 2, 1 ; 1-x\right)$.
An interesting feature of these functions is that while $F_1(0)=1$ and it expands about $x=0$ into a standard Taylor series, the function $F_2(x)$ behaves as $\ln(x)$ close to $x=0$.
It is straightforward to check that the conformal block $F_2(x)$ is then compatible with the OPE
\be \sigma(z) \sigma(0) \sim z^{\frac 1 4} \left(  \ln z + : \bar \psi \psi :(0) + \dots \right).
\ee
Thus, the logarithmic operator appears on the right hand side of the OPE $\sigma(z) \sigma(0)$ and is responsible for the logarithmic behavior of $F_2(x)$. It can be compared with \rfs{eq:opel} to confirm that $:\bar \psi \psi$ clearly is a logarithmic partner of an identity operator.

The reason for the appearance of the logarithmic operators in the expansion of $\sigma(z)$ with itself can be read off the Kac table, in accordance with the remarks at the end of Sec.~\ref{sec:log} following
\rfs{eq:last}. According to the standard rules
of the OPE of an operator with itself, the expansion of the operator $(1,2)$ with itself can give either the $(1,1)$ operator or the $(1,3)$ operator. In the Kac table both of
these have dimension zero. One might conclude that this problem simply has two dimension zero operators, $(1,1)$ and $(1,3)$. However, the matching of their dimensions implies that
the corresponding differential equations have two solutions which both have the same power law singularity at the origin $F(x) \sim x^{\alpha}$ (the power reflects the dimensions of the operators). It is well known
in the theory of differential equations that under these conditions while the first solution goes as $F(x) \sim x^{\alpha}$, the second solution must go as $F(x) \sim x^{\alpha} \ln x$, 
due to the ``collision of dimensions". 

The bottom line, confirming the remarks at the end of Sec.~\ref{sec:log}, the logarithmic operators naturally appear in the OPE of two degenerate operators  from the Kac table such that, when their expansion is computed according to the rules
of the Kac table, it includes on the right hand side more than one operator with the same dimension. 

Moreover, it is also well known in the theory of differential equations that if an equation appears to have two solutions, one going as $F_1(x) \sim x^{\alpha}$ at small $x$ and another one going as
$F_2(x) \sim x^{\alpha + n}$, with $n$ some positive integer, then the solution $F_1(x)$ must have in its expansion in powers of $x$ the term $F_1(x) \sim x^{\alpha} + \dots + c^{\alpha+n} \ln x + \dots$. 
It follows from this that 
 the logarithms also appear if the dimensions of the operators
 on the right hand side of the OPEs of primary degenerate operators are different by an integer. This will play a crucial role in the next section. 

\section{Logarithmic operators at $c=0$}
\label{sec:zero}
\subsection{Logarithmic partner of the stress-energy tensor}
Theories with vanishing central charge play a crucial role in studying models of condensed matter physics which involve quenched disorder. Therefore, these theories have been subject of
extensive research, and our inability of solving most of these models remain a stumbling block on the way to the theory of critical points with quenched disorder.

An interesting feature of the theories with the central charge $c=0$ is that they must involve an operator with dimension $2$ other than stress energy tensor such that $L_2$, when applied to this operator,
gives a nonzero constant. Indeed, a primary operator with nonvanishing two point function in any conformal field theory with $c \not = 0$ must have an OPE with itself which goes as
\be  \label{eq:cat} A(z) A(0) = \frac{1}{z^{2h}} \left (1 + \frac{2h}{c} z^2 T(0) + \dots \right).
\ee 
A direct limit $c \rightarrow 0$ in this equation is not possible, resulting in a ``$c \rightarrow 0$ catastrophe''. To resolve it, we recall the origin of the coefficient $2h/c$ in \rfs{eq:cat}.  If one applies $L_2$ to both sides of this equation, on the one hand one finds
\be \label{eq:oneway} \frac{2h}{c} z^{2-2h} L_2 T(0)= h z^{2-2h},
\ee since $L_2 T(0)=c/2$. On the other hand, we can employ  the commutation relations \rfs{eq:st} to find
\begin{eqnarray} \label{eq:secondway} && L_2 A(z) A(0) = \left[ L_2, A(z) \right] A(0) = \left[ z^{3} \pp{z} + 3 z^2 h \right] A(z) A(0) \approx \cr && \left[ z^{3} \pp{z} + 3 z^2  h \right] \frac{1}{z^{2 h}}=h z^{2-2h}.
\end{eqnarray}
Equating this with \rfs{eq:oneway} we find that the choice of the coefficient $2h/c$ in \rfs{eq:cat} is the only one which makes these two calculations to give the same result. 

Further comparison of Eqs.~\rf{eq:oneway} and \rf{eq:secondway} shows that the only way to make these compatible at $c=0$ is to assume the existence of another dimension $2$ operator, which we call $t(z)$, such that \be L_2 t(0) = b,\ee where $b$ is some nonzero coefficient \cite{Gurarie1999}. Then one can suppose that
\be \label{eq:cat1} A(z) A(0) \approx \frac{1}{z^{2 h}} \left(1+\frac h b z^2 t(z) + \dots \right),
\ee replacing the OPE \rfs{eq:cat}. This OPE can be verified to be consistent under the application of $L_2$ to both of its sides. 
Further requirements on $t(z)$ include
\be L_n t(0) = 0, \ n>2 \ {\rm and} \ n=1.
\ee
Other than that, $t(z)$ remains arbitrary. 

Some conformal field theories at $c=0$ may have $t(z)$ which is a quasiprimary field, that is,
\be L_2 t(0)=b, \ L_0 t(0)=2 t(0),
\ee
or equivalently
\be \label{eq:simpleTt} T(z) t(w) = \frac{b}{(z-w)^4} + \frac{2 t(w)} {(z-w)^2}+\frac{t'(w)}{z-w}+\dots.
\ee
A good example of this is a direct sum of two conformal field theories with central charges $c_1=c$ and $c_2=-c$, whose total central charge is zero. Then one can argue that $T=T_1+T_2$ and $t=T_1-T_2$, satisfying \rfs{eq:simpleTt} with $b=c$. 

However an intriguing possibility is that $t(z)$ is a logarithmic partner of  the stress-energy tensor $T(z)$. This is indeed possible, if one takes into account that due to the fact that $c=0$, $T(z)$ has a vanishing norm,
\be \label{eq:stressnorm} \VEV{T(z) \, T(w)} =0.
\ee
This is compatible with the basic property of logarithmic partners \rfs{eq:logcorr}. Therefore, we can postulate the OPE generalizing \rfs{eq:simpleTt}
the OPE
\be \label{eq:Tt} T(z) t(w) = \frac{b}{(z-w)^4} + \frac{2 t(w) + T(w)}{(z-w)^2} + \frac{t'(w)}{z-w} + \dots.
\ee
It immediately leads, by \rfs{eq:logcorr}, to the following correlation functions
\begin{eqnarray} \label{eq:allcorr} && \VEV{T(z) T(w)} =0, \ \VEV{t(z) T(w)} = \VEV{T(z) t(w)} = \frac{b}{(z-w)^4}, \cr && \VEV{t(z) t(w)} = -2 b \frac{ \ln (z-w) }{(z-w)^4}.
\end{eqnarray}
At the same time, \rfs{eq:cat1} gets further modified, to give
\be \label{eq:logope} A(z) A(0) = \frac{1}{z^{2 \Delta}} \left( 1+ \frac h b \left( t(z) + T(z) \ln z \right) + \dots \right),
\ee 
as is clear if one examines \rfs{eq:opel}. 

Further analysis of the OPE between two operators $t$ leads to the following expressions, constructed just like elsewhere in this paper by the consistent applications of $L_n$ with $n \ge 0$ to the both
sides of this equation, following the method described after \rfs{eq:st}
\begin{eqnarray} \label{eq:tt} t(z) t(0) &=& \frac{-2b \ln (z)}{z^4} +\frac{t(0) \left[ 1- \ln (z) \right] - T(0) \left[ \ln(z) + 2 \ln^2 (z) \right]}{z^2} +\cr
&& +\frac{t'(0) \left[ 1- 4 \ln(z) \right]-T'(0) \left[ \ln(z) + 2 \ln^2 (z) \right]}{2z} + \dots.
\end{eqnarray}

\subsection{Logarithmic algebra at $c=0$}
\label{sec:algebra}
It is well known that the OPE of the stress energy tensor with itself generates the Virasoro algebra. 
One could ask if the OPEs given by \rf{eq:Tt} and \rf{eq:tt} generate an extension of the Virasoro algebra which could then be used to construct
extended descendants of primary operators the way it is usually done with the Virasoro algebra. 

The full answer to this question is not known, but a partial answer can be given in the following way.
A primary operator $A(z)$, when contracted with the stress energy tensor generates an OPE which contains logarithms. The logarithms are not arbitrary, but rather can be encoded with the following expression
\be \label{eq:tA} t(z) A(0) = - T(z) A(0) \ln (z) + \sum_{n=0}^\infty A_n^t z^{n-2}.
\ee
In other words, it can be argued that the conformal invariance requires $\ln(z)$ to appear on the right hand side of this OPE, and the expression in front of the logarithm
must coincide with the OPE between $T(z) A(0)$, up to a sign. This can be verified by applying $L_0$ to both sides of this expression, in a way similar to 
the discussion following \rfs{eq:st}. $A_n^t$ are operators which can be deemed logarithmic descendants of the primary operator $A(0)$, leading to the definition
\be \ell_{-n} A = A^t_n.
\ee
Those can now be calculated using the usual expression
\be \ell_n = \oint \frac{dz}{2\pi i} \left( t(z) + T(z) \ln(z) \right) z^{n+1}.
\ee
Combined with the definition of the Virasoro generators
\be L_n = \oint \frac{dz}{2\pi i} T(z) z^{n+1},
\ee this leads to the following commutation relations, from the OPE \rfs{eq:Tt},
\be \label{eq:lL} \left[\ell_n, L_m \right] = (n-m) \ell_{n+m} - m L_{n+m} + \frac{b}{6} n (n^2-1) \delta_{n,-m}.
\ee
It would be natural to try to supplement these with the commutation relations $\left[ \ell_n, \ell_m \right]$. Unfortunately, these commutation relations are not known.
At the same time, the commutation relations \rfs{eq:lL} will be useful in the next subsections. Finally, we remark that the method used to construct \rfs{eq:lL} is somewhat reminiscent of the methods
used in Ref.~\cite{Fateev1985} to find the commutation relations of the parafermionic operators. 

\subsection{Kac table at $c=0$}
\label{sec:Kaczero}
A partner of the stress energy tensor $t(z)$ with $L_2 t(0) = b$ must exist in any $c=0$ theory, to avoid the ``$c \rightarrow 0$ catastrophe". However, $t(z)$ does not have to be logarithmic. Yet logarithmic 
$t(z)$ may occur in some circumstances.  The easiest example of this occurs for some of the operators from the $c=0$ Kac table. The table itself at vanishing central charge reads
\begin{center} \begin{tabular}{  l || c | c | c| c| r}  $n$ \textbackslash \, $m$ & 1 & 2 & 3 &4 & \dots \\ \hline \hline  \noalign{\smallskip} 1 & 0 & 0  & $\frac 1 3$ & 1 & \dots \\ 
\noalign{\smallskip} \hline \noalign{\smallskip} 2 &$ \frac 5 8 $ & $\frac 1 8$  &$ - \frac 1 {24} $& $\frac 1 8$ & \dots \\
\noalign{\smallskip} \hline \noalign{\smallskip} 3 & 2 & 1  & $\frac 1 3$ & 0 & \dots \\
\noalign{\smallskip} \hline \noalign{\smallskip} \dots & \dots & \dots & \dots & \dots & \dots
\end{tabular} 
\end{center}
Take the operator $(n,m)$=$(2,1)$ with conformal dimension $5/8$. This operator is degenerate on the second level, and its four point function satisfies a second order
differential equation. Not surprisingly in complete agreement with the observations at the end of Sec.~\ref{sec:minustwo} its solutions have logarithmic singularities. 
The origin of them is the fact that the operator at the position $(3,1)$ has dimension $2$, which differs by an integer from the dimension 0  of the identity operator at the position $(1,1)$. Both $(1,1)$
and $(3,1)$ appear in the OPE with $(2,1)$ with itself. 

The analysis of these singularities shows that they correspond to the OPE \rfs{eq:logope}. Moreover, they also fix the number $b$ in this OPE, which can be read off
the solution to this equation to be
$b=5/6$. 

Remarkably, in this realization of the $c=0$ theory, $t(z)$ is automatically logarithmic and $b$ is fixed.

On the other hand, one could study the third order equation satisfied by the operator $(1,3)$. That equation also has logarithmic solutions, corresponding to the OPE \rfs{eq:logope} but with $b=-5/8$. 

A natural question one could try to address is if one theory is allowed to have to separate $t(z)$, both obeying \rfs{eq:Tt} but with two distinct values of $b$. A simple calculation shows that if two such $t(z)$ exist, labelled by, say, $t_{b_1}(z)$ and $t_{b_2}(z)$, then the conformally invariant correlation function $\VEV{t_{b_1}(z_1) \, t_{b_2}(z_2)}$ would not be possible (it could not be invariant under translations, dilatations, and special conformal transformations). We are led to a startling conclusion that a degenerate operator $(2,1)$ and a degenerate operator $(1,3)$ cannot be simultaneously degenerate in the same conformal theory. 

On the other hand, a zero dimensional operator $(1,2)$, which we denote ${\cal O}(z)$, can resolve the ``$c \rightarrow 0$ catastrophe" in a different way. Its dimension is zero, and therefore the ratio of its dimension to the central charge is not necessarily infinity. In particular, one can check that the operator product expansion
\be \label{eq:OPEzero} {\cal O}(z_1) {\cal O}(z_2) \sim 1 + C z^2 \left( T(0) + \dots \right) + \dots,
\ee
which does not involve $t(z)$ and any logarithms. 
Here $C$ is a coefficient which cannot be fixed purely by conformal invariance, which can be interpreted as the limit of $2 h/c$ when both the dimension $h$ and the central charge $c$ are taken to zero. Interestingly, identifying $h$ as the dimension of the $(1,2)$ operator even at $c>0$, that is, for models controlled by $p=q+1$ at $q>2$ and taking the limit  $q \rightarrow 2$ equivalent to $c \rightarrow 0$, one can check that $C=1/5$. This same result can also be obtained from a slightly different point of view directly at $c=0$, as we will see below. 

\subsection{Logarithmic constraints on the Kac table}

Let us attempt to understand why an operator such as $(2,1)$ from the $c=0$ Kac table cannot be degenerate unless the parameter $b$ is equal to $5/6$. The operator $(2,1)$ is degenerate on the second level because
one can verify that 
\be \label{eq:descendant} \left( L_{-2} - \frac 2 3 L_{-1} L_{-1} \right) \left| (2,1) \right>
\ee
is a primary operator. That means, it is annihilated by any $L_n$ with $n>0$. In the usual unitary conformal field theory, this implies that this operator can be set to zero, because
it has zero norm. However, in our nonunitary theory, there are plenty of operators which have zero norm, such as the stress-energy tensor itself $T(z)$ as
discussed in \rfs{eq:stressnorm}, which are
nonetheless not zero. However, suppose that the operator \rfs{eq:descendant} is also annihilated by $\ell_n$. Then it may be more credible to declare that it has zero norm. For example, 
the stress energy tensor itself is obviously not annihilated by $t$, precisely because the correlation function of $t$ and $T$ \rfs{eq:allcorr} which is proportional to $b \not = 0$ is not zero. 

Applying $\ell_{1}$ to \rfs{eq:descendant} and using the commutation relations \rfs{eq:lL} we find
\be \label{eq:first} \left(  \ell_{-1} - \oh L_{-1} \right) \left| (2,1) \right>.
\ee
Applying $\ell_{2}$ to \rfs{eq:descendant} we find
\be \label{eq:second} \left( b - \frac 5 6 \right) \left| (2,1) \right>.
\ee
Applying $\ell_n$ with $n>2$ to \rfs{eq:descendant} automatically gives zero.

From \rfs{eq:second} it immediately follows that $b=5/6$. \rfs{eq:first} is a little more subtle. However, applying $L_{1}$ to it gives zero (applying $\ell_{1}$ to it is not
possible in view of the lack of knowledge of the commutation relations $\left[ \ell_n, \ell_m \right]$), which is compatible with setting it to zero. In deriving this it is necessary
to set $\ell_0 A(0) = 0$, which is natural in view of the fact that $T$ with an arbitrary coefficient can always be added to $t$ without modifying the logarithmic part of the OPEs where $t$ participates. Likewise, $L_n$ times
an arbitrary coefficient can always be added to $\ell_n$, leading to the possibility of setting $\ell_0 A(0)$ to zero as long as $L_0 A(0)$ is not zero. 

The end result, in order for the operator $(2,1)$ to be degenerate on the second level (and the appropriate differential equation for its correlation functions to hold), $b$ must be equal to $5/6$ as expected. 

By a systematic application of the commutation relations \rfs{eq:lL} it can be further established that $b=5/6$ for all the operators of the form $(n,1)$ and $(n,2)$ with $n>1$, while $b=-5/8$ for the operators of the form
$(1,m)$ with $m>2$ \cite{Momo}.

On the other hand, operators with vanishing dimension are exceptional  because even \rfs{eq:tA} is not valid for them. Rather, it has to be replaced by
\be t(z) {\cal O}(0) = - \left(1-\epsilon \right) T(z) {\cal O}(0) \ln z + \dots.
\ee
Here $\epsilon$ is a constant which cannot be fixed by conformal invariance alone.
Following the logic of Sec.~\ref{sec:algebra} we can derive a new form of the commutation relations between $\ell_n$ and $L_m$, at work only when applied to a zero dimensional operator. These now take the form
\be  \label{eq:lLnew} \left[\ell_n, L_m \right] = (n-m) \ell_{n+m} + \left( \epsilon-  m\right) L_{n+m} + \frac{b}{6} n (n^2-1) \delta_{n,-m}.
\ee
Moreover, $\ell_0$ when acting on ${\cal O}(0)$ no longer has to vanish. We denote
\be \ell_0 {\cal O}(0)= \Delta {\cal O}(0).
\ee
In particular, an operator $(1,2)$ which plays an important role in the theory of percolation is supposed to be degenerate on the second level, since it is supposed to have the null descendent 
\be \left( L_{-2} - \frac 3 2 L_{-1} L_{-1} \right) {\cal O}(0).
\ee
As in the previous example, we apply $\ell_{1}$ and $\ell_{2}$ to this using the new commutation relations \rfs{eq:lLnew}. Requiring that the result vanishes gives
\be \label{eq:condzero} b = 5 \Delta, \ \Delta= \frac{-5 +7 \epsilon}{12}.
\ee
These are the conditions various coefficients must satisfy in order for the operator $(1,2)$ to be degenerate on the second level and in order for its correlation functions to satisfy the appropriate correlation functions.

The first of these relations have an interesting and unexpected meaning. It can be verified that, first of all, 
\be \VEV{t(z) {\cal O}(w_1) {\cal O}(w_2)} = \Delta \left( \frac{ (w_1-w_1)}{(z-w_1) (z-w_2) } \right)^2.
\ee
The correlation function itself can be found by conformal invariance, using that $\VEV{T(z) {\cal O}(w_1) {\cal O}(w_2)}$ is zero by the standard Ward identity. The coefficient in front of it can be related to $\ell_0 {\cal O}(0)$ by taking the limit $z \rightarrow w_1$. Second, the correlation function 
$\VEV{ {\cal O}(z_1) {\cal O}(z_2) A(w_1) A(w_2) }$, where $A$ is any primary operator with nonzero dimension, can be studied in the limit where $z_1$ goes to $z_2$ first, using \rfs{eq:OPEzero}, and then 
in the limit where $w_1$ goes to $w_2$ first using \rfs{eq:logope}. Equating the results gives $C=\Delta/b$. From \rfs{eq:condzero} we now deduce that 
\be C = \frac 1 5,
\ee
a result previously derived at the end of Sec.~\ref{sec:Kaczero} 
using a completely different method which relied on taking $c>0$ and subsequently taking the limit $c \rightarrow 0$ in a certain way. 

As for the second relation given in 
\rfs{eq:condzero}, its meaning is far more obscure. For example, it was recently suggested that percolation theory must have $b=-5$ \cite{Saleur2012}. If so, this implies that $\Delta=-1$. Then this further implies by
\rfs{eq:condzero} that 
\be
\epsilon=-1,
\ee
implying the following OPEs for the $(1,2)$ operator
\be \label{eq:zerop} t(z) {\cal O}(0)= - 2 T(z) {\cal O}(0) \ln z + \dots,
\ee where $\dots$ stand for the terms not containing the logarithms.

The zero dimensional operator $(1,2)$ at $c=0$ plays a special role in the theory of percolation where its correlation function determines the crossing probability of the percolation clusters in finite geometry \cite{Cardy1992}. 
The differential equation its correlation function satisfies has been the tool used to compute the crossing probability analytically, with the result also having been verified numerically \cite{Pouliot1992}. We conclude therefore that 
the $(1,2)$ operator as understood in percolation must satisfy \rfs{eq:zerop}. 

Further development of this theory, and its applications to other disordered systems, is limited by the lack of the suitable approach to compute the correlators $\left[ \ell_n , \ell_m \right]$.

\subsection{Emergence of the supersymmetry multiplet}

One the existence of a logarithmic partner $t$ of the stress energy tensor $T$ is postulated, two more dimension 2 primary operators appear naturally. Take the OPE between two operators $t(z)$ and $t(0)$, \rfs{eq:tt}. Take $z$ around zero in the counter-clock-wise direction. Under this operation the logarithm goes into itself plus a constant $2\pi i$. That means, a piece is added to the OPE which starts with $- 4\pi i b/z^4$ (and continues as an expansion in powers of $z$ and $\ln z$). That piece itself must be an OPE between two primary dimension two operators with nonvanishing two point function, to prevent the violation of conformal invariance (that is, as always applying $L_n$ to both sides of the OPE just as in the discussion following \rfs{eq:st} must lead to the same result). 

Let us assume that these operators are fermionic. Then they must have the following OPE
\be \xi(z) \bar \xi(0) = \alpha T(z) T(0) + \frac{b}{2z^4} +\frac{t(0) +T(0) \ln z}{z^2} + \dots.
\ee
Here the part of this OPE beginning from the $b$-term is simply the general property \rfs{eq:logope} for $h=2$, multiplied by $b/2$ as a normalization, and the term proportional to arbitrary (so far) coefficient $\alpha$ reflects the fact that an OPE between $T(z)$ and $T(0)$ can always be added on the right hand side without violating conformal invariance. 

Now consider the combination
\be \label{eq:supplement} \VEV{ \xi(z_1) \bar \xi(z_2) \bar \xi(z_3) \xi(z_4)} - \oh \VEV{T(z_1) T(z_2) \bar \xi(z_3) \xi(z_4)} \ln \frac{(z_1-z_2)(z_3-z_4)}{(z_1-z_3)(z_2-z_4)}.
\ee
As a function of $z_1$ it is a single valued, and indeed rational, function, as follows from the OPEs between $\bar \xi$ and $\xi$, as well as between $T$ and the primary fields $\xi$ and $\bar \xi$, and the OPE of $T$ with itself. It can then be reconstructed by its poles. Since we demand that $\xi$ and $\bar \xi$ are fermionic, we would like the function in \rfs{eq:supplement} to vanish as $z_1$ approaches $z_4$, for example. It turns out (can be verified by direct algebra following the reconstruction of the function in \rfs{eq:supplement}) that it vanishes only if $\alpha=1/8$. Therefore,
\be \label{eq:xixi} \xi(z) \bar \xi(0) = \frac{ T(z) T(0)} 8 + \frac{b}{2z^4} +\frac{t(0) +T(0) \ln z}{z^2} + \dots.
\ee
Now the correlation function between $t$, $\xi$ and $\bar \xi$ can be found from conformal invariance (that is, imposing the invariance under translations, dilatations, and special conformal transformations, just as in simpler cases considered in Sec.~\ref{sec:log}) and the compatibly with the OPE constructed in \rfs{eq:xixi} to be
\be \label{eq:txixi}  \VEV{t(z) \xi(w_1) \bar \xi(w_2)} =b \frac{\ln \left( \frac{w_1-w_2}{(z-w_1)(z-w_2)} \right)+ \frac 1 4}{(z-w_1)^2 (z-w_2)^2 (w_1-w_2)^2}.
\ee
The following OPE then follows from this correlation function
\be \label{eq:txi} t(z) \xi(0) = \frac{T(z) \xi(0)}{4} - T(z) \xi(0) \ln z +\frac{\xi'(0)}{2z} + \dots.
\ee
The term proportional to $\ln z$ is required by conformal invariance, as in discussion after \rfs{eq:tA}, but the term $T(z) \xi(0)$ can appear there as well without violating conformal invariance. The specific coefficient in front of it in \rfs{eq:txi} is just a direct consequence of $\alpha=1/8$ which in turn fixed the coefficient $1/4$ in the numerator of \rfs{eq:txixi}. A similar expression can be derived replacing $\xi$ by $\bar \xi$. 

\begin{figure}
\includegraphics[width=5cm]{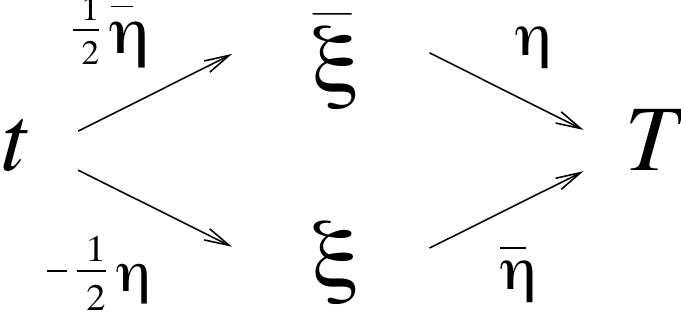}
\caption{\label{fig:mult}
Stress-energy multiplet.}
\end{figure}

Now a remarkable observation can be made which says that the OPEs $T(z) T(0)$, $t(z) T(0)$, $t(z) t(w)$, $t(z) \xi(w)$, $t(z) \bar \xi(w)$ and $\xi(z) \xi(w)$, given in Eqs.~\rf{eq:Tt}, \rf{eq:tt}, \rf{eq:txi}, and \rf{eq:xixi}
(the OPE between two stress energy tensors are the standard ones with $c=0$)  all transform according to the diagram shown in Fig.~\ref{fig:mult}.

This diagram shows that under acting on the fields by the fermionic generators of $U(1|1)$ unitary supergroup $\eta$ and $\bar \eta$, and assuming that $t$, $T$, $\xi$, $\bar \xi$ transform according to the Fig.~\ref{fig:mult}, the corresponding OPEs transform into each other. This shows that the resulting conformal field theory has a global $U(1|1)$ symmetry. 

Let us demonstrate this on the example of a $t(z) \bar \xi(0)$ OPE. 
Apply the operation $\eta$ to it according to the Fig.~\ref{fig:mult} (that is, transform $t$ and $\bar \xi$ according to the lines in the diagram corresponding to $\eta$ operation taking into account that $\eta(t \bar \xi) = \eta(t) \bar \xi+ t \eta(\xi)$). The result is
\be \label{eq:sss} - 2 \xi(z) \bar \xi(0) + t(z) T(0)= \frac{T(0)}{2 z^2} - \frac{2 T(0) \ln z}{z^2} + \dots,
\ee
where Eqs.~\rf{eq:xixi} and \rf{eq:Tt} were used. On the other hand, take the OPE $t(z) \bar \xi(w)$, which is the one given in \rfs{eq:txi} with $\xi$ replaced by $\bar \xi$, implying
\be t(z) \bar \xi(0) = \frac{\bar \xi(0)}{2 z^2} - \frac{2 \bar \xi(0) \ln(z)}{z^2} + \dots.
\ee
Now apply the operation $\eta$ according
to the Fig.~\ref{fig:mult} directly to its right hand side. The result coincides with \rfs{eq:sss}. That shows that this particular OPE correctly transforms under the supergroup generator $\eta$. Likewise, the invariance of other OPEs can also be verified. 

This is rather remarkable because various problems of a quantum particle moving in a presence of a spatially random potential whose critical points are expected to be described by $c=0$ conformal field theories map into field theories invariant under such super-rotations  (see \cite{KoganBook} for an extended discussion of this, or \cite{EfetovBook} for an introduction to the supersymmetry method in quantum problems with disorder). 
Furthermore, problems such as percolation and polymers, expected to be described by $c=0$ conformal field theory, also have supersymmetric formulations \cite{Parisi1980,Gruzberg1999}. 


We see that  the fields in all the $c=0$ theories with the logarithmic $t$ are automatically organized in supersymmetry multiplets, making it very natural to conjecture that among these theories are the solutions to these longstanding problems involving quenched disorder, as well as the problem of two dimensional percolation and self avoiding random walks. Yet unfortunately no complete classification of the CFTs with $c=0$ and logarithmic $t$ exists, and that could be the reason why all these problems remain unsolved (as previously discussed, two dimensional self avoiding random walks and percolation are partially understood, but it's fair to say that their complete critical theory is also lacking). 

\section{Conclusion and outlook}

Logarithmic conformal field theory turned out to be a rich subject whose study is still ongoing. While various mathematical aspects of logarithmic operators still need to be clarified, from the point of view of applications to disordered systems further development of the theory of logarithmic partner of the stress-energy tensor is imperative if these types of conformal field theories ever lead to the answers of some of the longstanding problems at vanishing central charge. 

One particular approach one could adopt which has not been considered here is to look at how the holomorphic and antiholomorphic sectors in the logarithmic theory are glued together. At $c=-2$, this is relatively easy. One could verify, for example, that in order to have single valued correlation functions, the operators $\sigma(z, \bar z)$ (whose holomorphic part was introduced after \rfs{eq:anti}) must have the OPE
\be \sigma(z, \bar z) \sigma(0,0)= (z \bar z)^{1/4} \left( \Xi + \ln (z \bar z) \right) + \dots,
\ee
where $\Xi$ is the zero dimensional logarithmic operator satisfying
\be L_0 \Xi = 1, \ \bar L_0 \Xi = 1.
\ee
It is somewhat more involved to understand how the holomorphic and antiholomorphic sectors at $c=-2$ are glued together. Part of the difficulty stems from the fact that from the point of view of supermultiplets, $t$ is not just a holomorphic field, but a two by two tensor, a partner of the stress-energy tensor $T_{\mu \nu}$ \cite{Gurarie1999}. Therefore, both $t$ and $\bar t$ must exist, and each may have an appropriate $z$ and $\bar z$ dependence. More on this can be found in Ref.~\cite{Cardy2013}. Ultimately resolving this difficulty may require deeper understanding of the relationship between the bulk and boundary logarithmic conformal field theory \cite{Saleur2012}. 

At the same time a somewhat different approach to this problem was pursued recently in a number of publications where instead of the continuum theory solvable lattice models are constructed whose continuum limit is believed to be described by a logarithmic conformal field theory \cite{Read2007,Vasseur2013}. 

It is hard to tell which of these approaches eventually lead to the answer we seek, but hopefully we will not have to wait long to find out.

\ack

The author would like to express his gratitude to many people for insightful discussions concerning logarithmic operators, but would like to especially thank A. Polyakov, A. Tsvelik, C. Nayak and M. Flohr, A. W. W. Ludwig, and J. Cardy. Sec.~\ref{sec:log} reflects some of the early discussions the author had with A. Polyakov; Sec.~\ref{sec:minustwo} is mostly based on the work with C. Nayak and M. Flohr \cite{Gurarie1997}; and Sec.~\ref{sec:zero} stems from a long-term collaboration with A. W. W. Ludwig whose results were published in Refs.~\cite{Gurarie2002,KoganBook}. 

The author is also grateful to NSF for support via grant no. DMR-1205303.

\vskip 1cm
\noindent {\bf References}
\vskip .5cm
\bibliography{LogCFT}
\bibliographystyle{unsrt}

\end{document}